\documentclass[aps,twocolumn,pre]{revtex4}
\usepackage{graphicx}
\usepackage{latexsym}
\usepackage{amsmath}
\usepackage{amsfonts}
\usepackage{amssymb}
\newcommand{\be}{\begin{equation}}
\newcommand{\ee}{\end{equation}}
\newcommand{\bea}{\begin{eqnarray}}
\newcommand{\eea}{\end{eqnarray}}
\begin{document}
\title{Heteropolymer Sequence Design and Preferential Solvation of Hydrophilic Monomers:
One More Application of Random Energy Model}
\author{Longhua Hu and Alexander Y. Grosberg}
\affiliation{Department of Physics, University of Minnesota, 116
Church Street SE, Minneapolis, MN 55455, USA}

\begin{abstract}

In this paper, we study the role of surface of the globule and the
role of interactions with the solvent for designed sequence
heteropolymers using random energy model (REM). We investigate the
ground state energy and surface monomer composition distribution. By
comparing the freezing transition in random and designed sequence
heteropolymers, we discuss the effects of design. Based on our
results, we are able to show under which conditions solvation effect
improves the quality of sequence design. Finally, we study sequence
space entropy and discuss the number of available sequences as a
function of imposed requirements for the design quality.

\end{abstract}

\maketitle

\section{Introduction}

The simplest concept taught to students about protein structure is
that hydrophobic monomers are mostly inside water-soluble globules,
while hydrophilic monomers are mostly on the surface. This beautiful
idea was around for over half a century \cite{Bresler1, Bresler2},
everyone agrees that it represents a cornerstone for our
understanding of proteins \cite{Ptitsyn_Finkelstein} - and yet it is
somehow neglected in the most sophisticated theories of
heteropolymers with quenched sequences.  Here, we have in mind the
train of works which started from pioneering contributions by
Bryngelson and Wolynes \cite{BW} and by Shakhnovich and Gutin
\cite{SG_IndependentInt}.  The insight of the former authors
\cite{BW} was to recognize the deep connection between protein field
and that of spin glasses and to apply the Random Energy Model
developed by Derrida \cite{REM_Derrida}; the contribution of the
latter \cite{SG_IndependentInt} was to actually derive the REM
approximation for a consistent microscopic model of a heteropolymer
with independent random interactions. By now, it is understood that
REM is a well controlled mean field approximation for the large
compact heteropolymer \cite{Sfatos}. The important part of
heteropolymer theory was also the idea of sequence design, which was
used both to better model proteins and to test heteropolymer
properties in general \cite{Shakhnovich_review_2006}.

What is important to emphasize is that heteropolymer freezing and
sequence design theories operate within the so-called volume
approximation, neglecting surface terms in energy. Our goal in the
present work is to investigate, for the simplest tractable model,
the interplay heteropolymer freezing and sequence design with
preferential solvation of some monomer species on the surface of
the globule.  In fact, even for random sequences preferential
solvation was not included in REM-based heteropolymer theory until
very recently \cite{Solvation_random}. Our work is ideologically a sequel to
the paper \cite{Solvation_random}, and we will use the ideas of that work.

In the recent series of works \cite{Khokhlov_PRL, Khokhlov_COSSMS}, sequence design was
discussed (under a different name of ``coloring'') in a slightly
different prospective, with an eye on chemically preparing
protein-like copolymers.  The solvation effect was given a very
prominent role in these works.  One of the goals of our work is to
make a closer link between various implementations of the sequence
design paradigm.

The paper is organized as follows. First we will introduce the
solvation model (section \ref{sec:solvation} and Fig.
\ref{fig:model_cartoon}). Then we will talk about sequence design
technique and how it is affected by the solvation (section
\ref{sec:design}). Surface monomer composition distribution is
obtained in section \ref{sec:design_by_solvation}. Our major results
are summarized in the phase diagram of the system, sketched in the
Fig. \ref{fig:phase_diagram_2}. Finally, we discuss in section
\ref{sec:sequence_entropy} the availability of sequences as a
function of their quality characterized by the energy gap between
their ground state and the majority of other states.

\section{The model}

\subsection{Energy: bulk and surface terms}\label{sec:solvation}

In our model, each monomer is assigned a quenched random variable
$\sigma$, which represents its monomer type.  For the random
sequence, we assume that $\sigma$ for each monomer is drawn from
some probability distribution $p\left(\sigma\right)$. For
simplicity, we restrict $p\left(\sigma\right)$ to have zero
average $\int\sigma p\left(\sigma\right)d\sigma=0$ and unit
variance $\int \sigma^2 p\left(\sigma\right)d\sigma=1$. There are
20 possible values of $\sigma$ for natural proteins since the
number of amino acids is 20. Theoretically it is convenient to
consider a continuous distribution of $\sigma$ or a discrete
distribution of just 2 monomer types.

In our model, energy of the system consists of contributions from
direct contact between monomers and of the contribution of
contacts between monomers and solvent.  The former, contact
energy, has a ``homopolymeric'' strong average attraction part
$\overline{B}$ independent on monomer type, and a
``heteropolymeric'' contribution $- \delta \!  B \sigma_i
\sigma_j$ with amplitude $ \delta \!  B$.  We assume that
$\overline{B}$ is sufficiently large, such that the globule is
quite dense, and the contacts with the solvent take place only on
the surface of the globule.  We mostly look at the case $ \delta
\! B>0$, such that similar monomers attract each other.  We assume
that each contact with the solvent provides energy $-\Gamma
\sigma_i$.  Thus, since $\Gamma > 0$, monomers with $\sigma>0$ are
hydrophilic, while those with $\sigma<0$ are hydrophobic.  Thus,
Hamiltonian of our model depends on the sequence, presented by
${\rm seq} = \{ \sigma_i \}$, and conformation, specified by
positions of all monomers ${\rm conf} = \{ \mathbf{r}_i \}$.  The
Hamiltonian reads
\begin{equation} H \left({\rm seq},{\rm conf}\right) = \sum_{i<j}^N
\left(\overline B -  \delta \!  B \sigma_i \sigma_j \right)
\Delta_{ij} -  \Gamma \sum_{i \in G}^K \sigma_i \ .
 \label{eq:Hamiltonian}
\ee
Here $G$ is the set of $K \sim N^{2/3}$ monomers located on the
surface and, therefore, exposed to the solvent.  $\Delta_{ij}
\equiv \Delta \left(  \mathbf{r}_i- \mathbf{r}_j  \right)$ is
contact map defined as:
\begin{equation}
\Delta_{ij} = \begin{cases}
1 & \text{$\mathbf{r}_i$ and $\mathbf{r}_j$ are nearest neighbors} \\
0 & \text{otherwise}
\end{cases} \ .
\end{equation}
As we mentioned, we only consider the most compact conformations,
and we assume there are $Q$ contacts in every conformation, so
$\sum_{i<j}^N \Delta_{ij} = Q$. For compact globule, $Q \sim N$.
For simplicity, we just use $N$ to denote the number of contacts
in a conformation.

Apart from the surface term, this model is equivalent to the one
considered earlier, e.g., \cite{BJ}.  A cartoon of the model is
sketched in Fig. \ref{fig:model_cartoon}.

\begin{figure}
\centerline{\scalebox{0.4}{\includegraphics{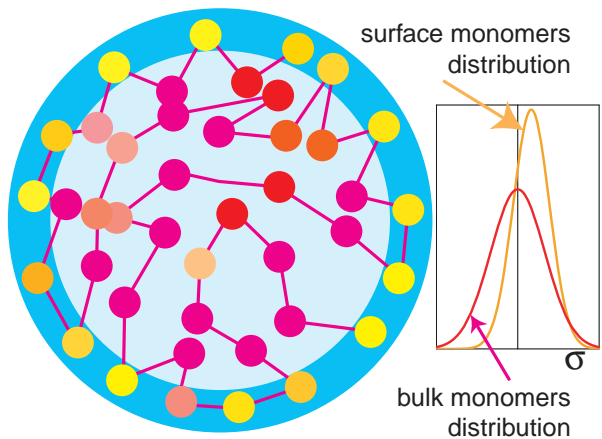}}}
\caption{(Color online) Illustration of the model. Monomers are
connected by covalent bonds, and monomer type is presented by the
shade. Surface is enriched with hydrophilic monomers, such that
the distribution of surface monomers  $f(\sigma)$ is shifted
compared with the bare distribution $p(\sigma)$.}\label{fig:model_cartoon}
\end{figure}

Polymer chain with Hamiltonian (\ref{eq:Hamiltonian}), with
quenched sequence and in the statistical equilibrium in terms of
conformation, will exhibit some preference of hydrophilic monomers
towards the surface - as long as $\Gamma >0$.  The way to
characterize this quantitatively is to address the statistics of
$\sigma$ values for surface exposed monomers.  Namely, we will be
interested in distribution $f\left( \sigma \right)$ of surface
monomers.  We expect this distribution to be different from the
bare distribution of all monomers $p \left( \sigma \right)$.
Qualitatively, this is illustrated in the inset of Fig. \ref{fig:model_cartoon}.

Of course, the effect of surface exposure to the solvent depends
on the sequence.  In general, hydrophilic effect adds frustration
to the system.  Indeed, placing a certain monomer on the surface
necessitates placing its sequence neighbors close to the surface,
while their identity, or their $\sigma$-values, might imply
energetic preference for the interior region of the globule.  To
address this delicate sequence dependence, we will look at
designed sequences.

\subsection{Sequence design}\label{sec:design}

Random sequence can be generated by a suitable Poisson process,
i.e., by the probability distribution
\begin{equation} P_{{\rm seq}}^{ \left(0 \right)} = \prod_{i=1}^N
p\left(\sigma_i\right) \ . \label{eq:propensities} \end{equation}
By sequence design, we want to bias the sequence probabilities in
a controlled fashion.  This can be done in the following way.

The sequence design procedure starts from the choice of the target
conformation which we will denote ${\star}$.  In our consideration
here, ${\star}$ might be any compact conformation, in other words,
we ignore the difference of designability for possible target
conformations \cite{Wingreen} - simply because designability and
surface exposure are two independent effects and we do not want to
further complicate our work by accounting for designability. There
is no doubt that designability along with surface effects must be
incorporated into the complete theory.  Given a target
conformation ${\star}$, we will consider a statistical Gibbs
ensemble in which conformation $\{ \mathbf{r}_i \} = {\star}$ is
quenched, while the sequence $\{ \sigma_i \}$ is annealed and
comes to thermodynamic equilibrium at the design temperature
$T_{d}$, which is not necessarily equal to the real temperature
$T$.  More specifically, we use the canonical sequence design
scheme \cite{BJ}, in the sense that it is generated by the
canonical ensemble of annealed sequences.  This results in the
following probability distribution of sequences:
\begin{equation}
P_{{\rm seq}}^{\star}=P_{{\rm seq}}^{ \left(0 \right)} \frac{ \exp
\left[-H^d \left({ \rm seq},{\star} \right)/T_d \right]}{ \sum_{{
\rm seq}^{\prime}}P_{{ \rm seq}^{\prime}}^{ \left(0 \right)} \exp
\left[-H^d \left({\rm seq}^{\prime},{\star} \right) /T_d \right]}
\ , \label{eq:probability_sequences}
\end{equation}
where the denominator ensures normalization.  Of course, this is
just the scheme, the real features of the ensemble of designed
sequences are controlled by the design Hamiltonian, $H^d \left({
\rm seq},{\star} \right)$.  It might be the same Hamiltonian as
(\ref{eq:Hamiltonian}), but this is not at all necessary.
Moreover, it is useful to explore the more general situation in
which $H^{d} \neq H$.  We will use design Hamiltonian of the same
functional form as (\ref{eq:Hamiltonian}), but with the different
parameters $\delta \! B^d$, and $\Gamma^d$ ($\overline{B^d}$,
although formally included for symmetry, does not play any role in
sequence design, because this term in energy is
sequence-independent):
\begin{equation} H^d \left({\rm seq},{\star}\right) = \sum_{i<j}^N \left
(\overline{ B^d} - \delta \! B^d \sigma_i \sigma_j \right)
\Delta_{ij} -  \Gamma^d \sum_{i \in G^{\star}}^K \sigma_i \ .
\label{eq:design_Hamiltonian} \end{equation}
From this point of view, $T_d$ is just a parameter which controls
the quality of design: when $T_d \to \infty$, ensemble of designed
sequences is essentially the same as random, while at lower $T_d$
designed sequences are statistically strongly biased by
optimization the design energy.

We work on artificial model proteins design. In vivo De Novo design of
globular protein has long been successfully carried out (see,
for example \cite{Degrado} for a four-helix protein design from first-principle).
The off-lattice model sequence design works (\cite{Broglia1, Broglia2}) on a
60-residue SH3 protein domain showed that the folding quality of designed sequences
vary for this small protein.

\subsection{High $T_d$ expansion of the
sequence-averaged target state energy and Random Energy Model}\label{sec:high_expansion}

First let us look at the energy of target conformation ${\star}$,
averaged over the ensemble of designed sequences:
\begin{equation} \langle E_{{\star}}\left({\rm seq}\right)\rangle = \sum_{{\rm
seq}} P_{{\rm seq}}^{\star} H \left({\rm seq},{\star} \right) \ .
\ee
As in \cite{BJ}, we can evaluate this to the lowest order in
high $T_d$ expansion:
\begin{eqnarray}
\langle E_{\star} \left({\rm seq}\right) \rangle & = & \langle H
\rangle + \frac{1}{T_d} \left(\langle H^d H \rangle - \langle H^d
\rangle \langle H \rangle \right)
\nonumber \\
& = & N \overline B - \frac{1}{T_d} \left(N \delta \!  B \delta \!
B^d + K \Gamma \Gamma^d \right) \ .
\label{eq:averaged_ground_state_energy}
\end{eqnarray}
We see that sequence design, on average, leads to the lowering of
the target state energy. The only novelty, albeit quite trivial,
compared at the volume approximation \cite{BJ}, is that both
contact energy and surface energy terms contribute to the target
energy decrease.

A'priori, one could think that the design lowers not only the
target state energy, but also energies of some other states -
particularly, those similar to the target state.  It is well
known, however, that because of the geometry of compact
conformations there are not many sufficiently similar compact
conformations and, therefore, the statistics of energies of other
conformations, to a good approximation, remains unaffected by the
design (see \cite{RMP} for further more detailed discussion of
this point).  This approximation is equivalent to  REM.  We will
work within this REM approximation.

\section{Monomer distributions inside and on the surface}

\subsection{Design by solvation}\label{sec:design_by_solvation}

As stated above, our major goal is to address surface effects in
terms of the distribution of $\sigma$ values among surface
monomers. With the designed sequences, we can first ask - what is
the distribution of surface monomers $p^{\star}(\sigma)$ just in
the design state, when conformation is frozen?  This is of course
fundamentally simple question, because the statistical mechanics
of sequences at quenched conformation is not frustrated
\cite{RMP}.  Basically, what one has to do is to take probability
distribution of sequences (\ref{eq:probability_sequences}) and to
integrate out all spin variables $\{ \sigma _i\}$ except surface
monomers, i.e., all $i$ except those belonging to the target
conformation surface set $G^{\star}$.

As a warm-up, let us perform this procedure for the simple yet
important special case of design Hamiltonian, in which we set $
\delta \! B^d=0$.  In this case, design affects only surface
monomers.  The probability distribution of sequences is
tremendously simplified, it can be factorized
\begin{eqnarray}  P_{{\rm seq}}^{\star} & = & P_{{\rm seq}}^{ \left(0 \right)}
\frac{ \exp \left[\left( \Gamma^d /T_d \right) \sum_{i \in
G^{\star}}^K \sigma_i  \right]}{ \sum_{{ \rm seq}^{\prime}}P_{{
\rm seq}^{\prime}}^{ \left(0 \right)} \exp \left[\left( \Gamma^d
/ T_d \right) \sum_{i \in G^{\star}}^K \sigma_i^{\prime}  \right]}
\nonumber \\
& = & \prod_{j \notin  G^{\star}} p(\sigma_j) \prod_{i \in
G^{\star}} p^{\star} (\sigma_i) \ ,
\label{eq:probability_sequences_deltaBdequals0} \end{eqnarray}
which means that the ensemble of monomers in the bulk of the
globule remains unaffected, distributed as $p(\sigma)$, while
every surface monomer, independently of others, is distributed as
\begin{equation}
p^{\star} \left( \sigma \right) = c p \left( \sigma \right) \exp
\left( \Gamma^d \sigma /T_d \right) \ , \label{eq:pstar}
\end{equation}
where $c$ is the normalization factor, $c=1 / \int p \left(
\sigma^{\prime} \right) \exp \left(\Gamma^d \sigma^{\prime}/T_d
\right) d\sigma^{\prime}$. Thus, for all monomers, including $N-K$
monomers that were in the bulk during the design process, and $K$
monomers that were on the surface, the overall distribution of
$\sigma$ reads
\begin{eqnarray}  p_{\rm tot} (\sigma) & = & \frac{(N-K) p(\sigma) + K p^{\star}
(\sigma)}{N} \nonumber \\ & = & p(\sigma) + \frac{K}{N} \left[
p^{\star} (\sigma) - p (\sigma) \right] \ . \label{eq:p_total1}
\end{eqnarray}
This result, Eq. (\ref{eq:pstar}), indicates that even this
simplified design procedure, with $\delta \! B^d =0$, favors the
hydrophilic monomers on the surface, because for hydrophilic
monomers with $\sigma>0$, we have $p^{\star} \left( \sigma \right)
> p \left( \sigma \right)$.  Compared with the bare monomer distribution
$p\left(\sigma\right)$, hydrophilic monomers have larger
probability to appear on the surface of target conformation.

The simplified $\delta \! B^d=0$ design scheme is reminiscent of
the method used in \cite{Khokhlov_PRL, Khokhlov_COSSMS}, in which all the surface
monomers of target conformation ($G^{\star}$ in our notation) are
made hydrophilic while all the monomers inside the globule are
hydrophobic. In our more general consideration, it is just more
probable but not necessary for the surface exposed monomers to
become hydrophilic during the design. The model of the works
\cite{Solvation_random} corresponds to the $T_d \to 0$ limit of our theory.

\subsection{Surface monomer distribution in the ground state}
\label{sec:distribution_derivation}

Let us continue examination of the simplified design scheme, with
$\delta \! B^d =0$, when only surface energy biases the choice of
sequences (design by solvation).  Our goal now is to find the
surface energy correction terms of the ground state energy and, as
the major step in this direction, we need to consider the surface
monomer distribution $f(\sigma)$ in the ground state. One should
realize that in the ground state, the set of surface exposed
monomers may or may not be similar to the set of monomers exposed
to the surface during the design; in other words, $f(\sigma)$
might be similar to $p^{\star} (\sigma)$, or might be quite
different from it. Therefore, there are two contributions to the
surface energy, one due to the monomers exposed to the surface,
and the other due to the fact that selection of surface monomers
affects the monomer composition left inside the globule.  To
express this quantitatively, we write for the arbitrary state the
equation similar to (\ref{eq:p_total1})
\begin{equation} p_{\rm tot} (\sigma) = p_{\rm in}(\sigma) + \frac{K}{N} \left[
f (\sigma) - p_{\rm in} (\sigma) \right] \ , \label{eq:p_total2}
\ee
where $p_{\rm in} (\sigma)$ is the distribution of monomers left
inside.  Comparing equation (\ref{eq:p_total1}) and
(\ref{eq:p_total2}), we find
\begin{equation} p_{\rm in} (\sigma) \simeq p(\sigma) + \frac{K}{N} \left[
p^{\star} (\sigma) - f(\sigma) \right] \ . \end{equation}
As everywhere, we neglect here the terms ${\cal O} \left((K/N)^2
\right)$.  Thus, we directly see already here how the deviation of
the state from the target state comes into play.

To compute $f(\sigma)$ for the ground state, we adapt for designed
sequences the procedure which was developed in the work
\cite{Solvation_random} for random sequence solvation. To make our work
self-contained, we briefly outline major steps.

We begin by constructing a separate REM, called sub-REM, for each
possible choice of surface monomers, $G$.  Indeed, for each $G$
there are still many conformations available.  The number of such
conformations is naturally written in the form $M_G = e^{Ns - K
\omega_G}$, where $s$ is conformational entropy per monomer in
volume approximation, and $\omega_G$ is entropy loss due to
confinement of some monomers on the surface.  Although $M_G$ is
much smaller than the total number of conformations, $M=e^{sN}$,
but the entropy loss caused by fixation of $G$ monomers on the
surface is only a surface effect ${\cal O}(K)$.  Following
\cite{Solvation_random}, we adopt a bold approximation that $\omega_G =
\overline{\omega}$ is independent on $G$; in this approximation,
counting all states shows that
${\overline \omega} = \ln \left( N e / K \right)$.

For each sub-REM, energies of all $M_G$ states are random in the
sense that they depend on random sequence realization, and it is
reasonable to assume \cite{Solvation_random} that these energies are
independent Gaussian variables, because each energy, according to
formula (\ref{eq:Hamiltonian}), has of order $N$ mutually
statistically independent bulk contributions and of order $K$
independent surface contributions.  To write down the resulting
Gaussian distribution, we should determine corresponding mean and
variance. The mean is found by averaging the bulk terms $\left(
\overline{B} - \delta \! B \sigma_i \sigma_j \right)$ over the
distribution $p_{\rm in} (\sigma)$ plus averaging the surface
terms $- \Gamma \sigma_i$ over the distribution $f(\sigma)$, and
the variance is similarly found by averaging the second moment.
This results finally in the following Gaussian distribution of
random energy:
\begin{equation}
w_G \left( E \right) \propto \exp \left[ -\frac{ \left[ E - \left(
N \overline B - K \Gamma \gamma_G \right) \right] ^2}{2 N \delta
\! B^2 + 2 K \delta \! B^2 \beta_G } \right] \ ,
\label{eq:Gaussian_distribution_of_levels}
\end{equation}
where
\begin{eqnarray}  \gamma_G & = & \int \sigma f \left(\sigma \right) d \sigma \ ,
\nonumber \\ \beta_G & =&  2 \int \sigma^2 \left[ p^{\star} \left(
\sigma \right) - f \left( \sigma \right) \right] d \sigma \ .
\label{eq:beta_definition} \end{eqnarray}
Notice that dependence on the surface monomer group $G$ is only through
the surface monomer distribution, and that the dependence on design is
due to the $p^{\star}(\sigma)$.

Every sub-REM has a certain ground state energy $E_g (G) = E_g\{f
(\sigma) \}$, which is just the lowest of $M_G$ random energies
drawn independently from the distribution
(\ref{eq:Gaussian_distribution_of_levels}). Energy $E_g\{f
(\sigma) \}$ is still a random variable, its probability
distribution can be found from the so-called extreme value
statistics \cite{Bouchaud_Mezard} (see also Appendix
\ref{sec:REM_distrib}):
\begin{equation}
{\cal W}_G \left( E \right) = \frac{1}{T_{\rm fr}} \exp \left[
\frac{\delta \! E_g}{T_{\rm fr}} - \exp \left[ \frac{\delta \! E_g
}{ T_{\rm fr} } \right] \right] \ ,
\label{eq:ground_state_distribution_in_the_group}
\end{equation}
where $T_{\rm fr} = \delta \! B / \sqrt{2 s}$ and $\delta \! E_g =
E - E_g^{\rm typ}\{f (\sigma) \}$ is the deviation of ground state
energy from its most probable (typical) value, which includes both
volume and surface contributions:
\begin{eqnarray}  E_g^{\rm typ}\{f (\sigma) \} & = & N \left( \overline B - 2 s
T_{\rm fr} \right) \nonumber \\
& + & K \left( \overline \omega T_{\rm fr}  - \Gamma \gamma_G + s
T_{\rm fr} \beta_G \right) \ .
\label{eq:typical_lowest_energy_in_the_group} \end{eqnarray}
Notice that the only dependence of probability distribution ${\cal W}_G$
on the surface monomers $G$ is hidden in $\gamma_G$ and $\beta_G$
inside the most probable energy $E_g^{\rm typ}\{f (\sigma) \}$,
and the dependence on design is also there inside $\beta_G$ (see
formula (\ref{eq:beta_definition}).  We also mention that $T_{\rm
fr} = \delta \! B / \sqrt{2 s}$ appearing here as a parameter of
the ground state distribution happens to have its physical meaning
- it is volume approximated freezing temperature of the random
sequence polymer \cite{Solvation_random}.

The probability to get ground state energy anywhere below its
typical most probable value
(\ref{eq:typical_lowest_energy_in_the_group}) is exponentially
small. However, we try exponentially many times - namely, we have
to choose the lowest among $e^{K {\overline \omega}} = (Ne / K)^K$
sub-REM ground states.  Therefore, we have a good chance to find
some particular sub-group $G$ with energy noticeably below typical
value (\ref{eq:typical_lowest_energy_in_the_group}). Essentially,
what we have to do now is to resort second time to the extreme
value statistics and find the expectation value of the lowest
among the sub-REM ground states.  It is convenient to perform this
operation in a slightly different, but equivalent form. Namely, we
note that the low energy tail of the ground state probability
distribution, (\ref{eq:ground_state_distribution_in_the_group}),
is exponential, and, therefore, it looks effectively like
Boltzmann distribution, with $T_{\rm fr}$ playing the role of
temperature.

It is useful to note here that treating the tail of the distribution
(\ref{eq:ground_state_distribution_in_the_group}) as effective
Boltzmann distribution with temperature $T_{\rm fr}$ is
reminiscent and essentially equivalent to the consideration given
in the book \cite{Ptitsyn_Finkelstein} and explaining the origin
of phenomenologically discovered quasi-Boltzmann distribution over
the ensemble of evolutionary selected proteins.

Returning to our argument, finding lowest among the $E_g^{\rm typ}$
of the sub-REMs is equivalent to minimizing the effective ``free
energy'', in which the effective entropy is given by the number of
ways to choose $K$ monomers with distribution $f (\sigma )$ from
$N$ monomers with distribution $p_{\rm tot} ( \sigma )$:
\begin{equation}
s\{f (\sigma) \}=-\int  p_{\rm tot} ( \sigma ) \left[ \phi \ln
\phi + \left( 1 - \phi \right) \ln \left( 1 - \phi \right) \right]
d \sigma \ ,
\end{equation}
where $\phi( \sigma ) = K f( \sigma )/ N p_{\rm tot}( \sigma )$
has the meaning of the fraction of monomers with type $\sigma$
that are exposed to the surface.  Including this effective
entropy, we have now the effective ``free energy''
\begin{equation}
E_g^{\rm typ} \{ f(\sigma ) \} - T_{\rm fr} N s  \{ f(\sigma ) \}
\ .
\end{equation}
We minimize this with respect to $f(\sigma )$, subject to
normalization condition $\int f(\sigma ) d \sigma = 1$ and obtain
\begin{equation} f(\sigma) = \frac{N}{K} \frac{p_{\rm tot} (\sigma)}{1 + \Lambda
e^{\eta_{\rm fr}(\sigma)}} \ , \label{eq:general_expression_for_f_simpler}
\end{equation}
where $\eta_{\rm fr}(\sigma)=2s\sigma^2 - \left(\Gamma / T_{\rm fr} \right) \sigma$,
and $\Lambda$ is the Lagrange multiplier which has to be
determined from the normalization condition $\int f(\sigma) d
\sigma = 1$.
Comparing this with the paper \cite{Solvation_random}, we see that the only role of design
in this case is the modification of monomer distribution: instead of bare distribution
$p(\sigma)$, we have now the modified one $p_{\rm tot}(\sigma)$.  Let us see what are
the consequence of this replacement.

Let us concentrate on the regime without the depletion effect, when $\phi(\sigma)\ll 1$
at all values of $\sigma$. This means that for any monomer type, only a small fraction
of it is solvated to the surface region. Under such assumption, $f(\sigma)$ can be
approximated as
\begin{equation}
f(\sigma)\propto
e^{-\eta_{\rm fr}(\sigma)}\left[1+\frac{K}{N}\left(c\exp\left(\frac{\Gamma^d}{T_d}
\sigma\right)-1\right)\right]p(\sigma) \  ,
\label{eq:surface_distribution}
\end{equation}
where we dropped for simplicity the $\sigma$-independent normalization factor.
To gain some insight, let us look at $f(\sigma)$ for a couple of simple examples of bare monomer probability
distributions.  Since real distribution involves a large number ($20$) of monomer species, we examine
two limits of two monomer species and of continuous Gaussian distribution.

\subsubsection{Example: bimodal distribution}
In the simplest black-and-white model \cite{Dill} two types of monomers, one hydrophilic and one hydrophobic,
appear with same probability:
\begin{equation}
p\left(\sigma\right)=\frac{1}{2}\left[ \delta \!
\left(\sigma+1\right)+\delta\left(\sigma-1\right)\right]\ .
\end{equation}
Simple calculation shows that
\begin{eqnarray}
f\left(\sigma\right) &\propto& \delta\left(\sigma+1\right)
e^{-\Gamma / T_{\rm fr}}\left[1-\frac{K}{N}
\tanh\left(\frac{\Gamma^d}{T_d}\right)\right]+\nonumber\\
&\hphantom{\propto}&\delta\left(\sigma-1\right)e^{ \Gamma / T_{\rm fr}}
\left[1+\frac{K}{N}\tanh\left(\frac{\Gamma^d}{T_d}\right)\right]\ .
\label{eq:surface_bimodal}
\end{eqnarray}
We see that there are two effects bringing hydrophilic monomers to the surface, that is,
increasing $f(+1)$ on the expense of decreasing $f(-1)$.  First effect is due to $\Gamma$ and
is measured by the ratio $\Gamma / T_{\rm fr}$.  This effect is present in random heteropolymer,
has nothing to do with design, and is simply energetic: since it is more favorable for the $\sigma > 0$
monomers to be on the surface, so the surface gets enriched with such monomers.  The second effect
is entirely due to design and it is governed by the design parameters $\Gamma^{d}/T_{d}$.  This effect
is washed way at large design temperature and it saturates at small $T_d$.  Notice that this design effect
is only a surface effect, its maximal possible role is proportional to $K/N$.  This is because the best one
can do with this type of design is to shift the monomeric composition by the amount about $K/N$.

\subsubsection{Example: Gaussian distribution}
The opposite limit is presented by
\begin{equation}
p\left(\sigma\right)=\frac{1}{\sqrt{2\pi}}\exp\left(-\frac{\sigma^2}{2}\right)\ .
\end{equation}
The corresponding surface monomer distribution is the following:
\begin{eqnarray}
f\left(\sigma\right)&\propto&\left[1+\frac{K}{N}\left[\exp\left(\frac{\Gamma^d\sigma}{T_d}-
\frac{1}{2}\left(\frac{\Gamma^d}{T_d}\right)^2\right)-1\right]\right]\nonumber\\
&\hphantom{=}&\times\exp\left(-\frac{1+4s}{2}\sigma^2+\frac{\Gamma}{
T_{\rm fr}}\sigma\right)\ .
\label{eq:surface_Gaussian}
\end{eqnarray}
In Fig. \ref{fig:gaussian_distribution}, we made a plot of an
example of $f(\sigma)$ as a function of $\sigma$ for the Gaussian
case. For comparison, random sequence solvation ($T_d\to\infty$) is
also included. It can be seen that design enriches surface with
hydrophilic monomers such that the distribution is shifted toward
hydrophilic region compared to no design case, which by itself is
already a shift relative to the bare distribution $p(\sigma)$.

\begin{figure}
\centerline{\scalebox{0.7}{\includegraphics{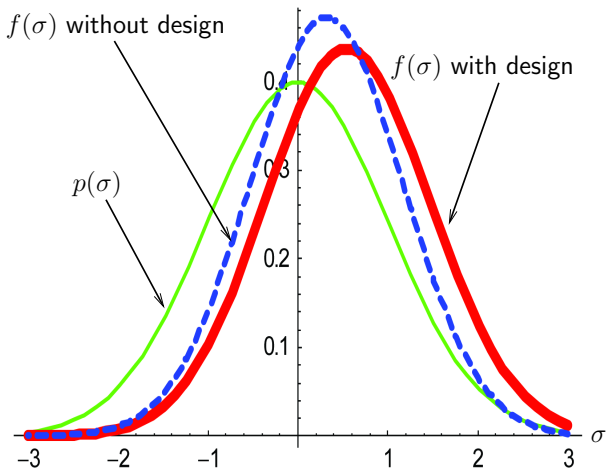}}}
\caption{(Color online). A comparison of original distribution and
surface monomer distribution with design, together with the case of
without design. Design favors the monomers with hydrophilic type.}
\label{fig:gaussian_distribution}
\end{figure}

\subsection{Sequence design by both solvation and monomer
contacts: mean field approximation}\label{sec:Design_Full}

In the preceding section, we considered sequences designed by the
effect of preferential solvation of certain monomer types under
the chain preparation conditions.  This corresponds to having only
the second term in the design Hamiltonian
(\ref{eq:design_Hamiltonian}) , or having $\delta \! B^{d} =0$. By
contrast, sequence design by monomer-monomer contacts, i.e. the
limit of $\Gamma^{d}=0$ and $\delta \! B^{d} \neq 0$ was
considered in the literature before (see, e.g., review article
\cite{RMP} and references therein).

In this section, we consider the general case, when both volume
and surface terms of the design Hamiltonian
(\ref{eq:design_Hamiltonian}) are present. To make the argument,
we resort to the mean-field approximation for the design system.
That means, we consider design by an effective field which couples
to the variables $\sigma$ and acts differently on surface and bulk
monomers.  Since design Hamiltonian (\ref{eq:design_Hamiltonian})
is quadratic in $\sigma$, the said ``design field'' is
proportional to $\overline{\sigma}$ - an average value of $\sigma$
defined self-consistently.  It is then easy to realize that this
field vanishes in the bulk, because in our model design does not
affect overall composition of the chain, and, therefore,
$\overline{\sigma}_{\rm bulk}=0$. Therefore, the mean field
approximated design Hamiltonian reads
\begin{equation}
H^d_{\rm mean \ field} \simeq -\left(\delta \! B^d z \overline{
\sigma}_{\rm surf} +\Gamma^d \right)\sum_{j \in G^{\star}}^K
\sigma_j \ . \label{eq:Design_hamiltonian_full}
\end{equation}
Here, $\overline{ \sigma}_{\rm surf}$ is the average $\sigma$ of
surface monomers (that is, of monomers which happen to be on the
surface during the design process) and $z$ is the coordination
number (the number of neighbors) for surface monomers.

Within the mean field approximation, the probability distribution
of designed sequences (\ref{eq:probability_sequences}) gets
factorized into independent distributions of all monomers, just
like in the $\delta \! B^d=0$ case.  Accordingly, we obtain the
distribution of surface monomers, similar to Eq. (\ref{eq:pstar}):
\begin{equation}
p^{\star}(\sigma) \propto p(\sigma) \exp\left[ \frac{ \delta \!
B^d z \overline{ \sigma}_{\rm surf} + \Gamma^d}{T_d}\sigma \right]
\ , \label{eq:Surface_target_full}
\end{equation}
where we dropped for brevity the $\sigma$-independent
normalization factor.  Now, the value of $\overline{ \sigma}_{\rm
surf}$ must be determined from the self-consistency condition
\begin{eqnarray} \overline{
\sigma}_{\rm surf} & = & \int \sigma p^{\star} (\sigma) d \sigma 
\nonumber \\ & = & \frac{\int \sigma p (\sigma) \exp\left[ \frac{
\delta \! B^d z \overline{ \sigma}_{\rm surf}+
\Gamma^d}{T_d}\sigma \right] d \sigma}{\int p (\sigma) \exp\left[
\frac{ \delta \! B^d z \overline{ \sigma}_{\rm surf}  + \Gamma^d
}{T_d}\sigma \right]  d \sigma} \ . \label{eq:selfconsistency}
\end{eqnarray}
To gain an insight into the properties of the latter equation, it is
useful to consider the examples of bimodal and Gaussian
distributions for $p(\sigma)$.  We do that a few lines below, in
section \ref{sec:self_consistency_by_examples}, but here we notice
that once $\overline{\sigma}_{\rm surf}$ is determined, the rest of the
analysis follows automatically along the lines of our previous
consideration in the section \ref{sec:distribution_derivation}.
Indeed, all we needed to know to implement the result
(\ref{eq:general_expression_for_f_simpler}) is the overall monomer
distribution $p_{\rm tot} (\sigma)$, which is known as soon as
$p^{\star}(\sigma)$ is determined (see Eq. (\ref{eq:p_total1})).
Therefore, we can directly use our results Eqs
(\ref{eq:surface_distribution}), (\ref{eq:surface_bimodal}),
(\ref{eq:surface_Gaussian}) with the replacement $\Gamma^d \to
\Gamma^d + \delta \! B^d z \overline{\sigma}_{\rm surf} \equiv
\Gamma^{\prime d}$.

With that in mind, let us return briefly to the determination of
$\overline{\sigma}_{\rm surf}$.

\subsection{Implementing the self-consistency condition}
\label{sec:self_consistency_by_examples}

\subsubsection{Example: bimodal distribution}

For bimodal $p(\sigma)$, Eq. (\ref{eq:selfconsistency}) becomes
\begin{equation} \overline{\sigma}_{\rm surf} = \tanh \left[ \frac{ \delta \!
B^d z \overline{ \sigma}_{\rm surf}+ \Gamma^d}{T_d} \right] \ .
\ee
At $\Gamma^d = 0$, this equation has non-trivial non-vanishing
solutions only if $\delta \! B^d z /T_d >1$, in which case there
are automatically two solutions of the opposite sign.  That means,
the non-zero $\overline{\sigma}_{\rm surf}$ in this case results
only from the spontaneous symmetry breaking, because without
$\Gamma^d$ the system has no preference for hydrophobic or
hydrophilic monomers dominating the surface.  The non-zero
$\Gamma^d >0$ breaks this symmetry and yields always one and only
one positive solution for $\overline{\sigma}_{\rm surf}$ (and
possibly two negative ones which we ignore because they have
higher free energy).

In this bimodal case, we have $\overline{\sigma}_{\rm surf} < 1$,
which means that in the replacement $\Gamma^d \to \Gamma^{\prime d}$, the solvation term
$\Gamma^d$ dominates if $\Gamma^d > \delta \! B^d z $.

\subsubsection{Example: Gaussian distribution}

For Gaussian $p(\sigma)$, Eq. (\ref{eq:selfconsistency}) is easily
explicitly resolved:
\begin{equation} \overline{\sigma}_{\rm surf} = \frac{\Gamma^d/T_d}{1 - z
\delta \! B^d /T_d}   \ . \end{equation}
This is usually very close to $\Gamma^d/T_d$, because (see next
section \ref{sec:phasediagram}) in most interesting regime close
to the triple point of the phase diagram, the denominator is
dominated by the unity.

\section{Free energy and phase diagram of designed polymers}
\label{sec:phasediagram}

\subsection{Preliminary remarks}

In this section, we will consider the possible phases of the
heteropolymer whose sequence is designed as discussed above.
 Similar problem in volume approximation is well known in the
literature (see, e.g., review \cite{RMP} and references therein).
Specifically, we will consider three phases and the transitions
between them.  We will summarize the relations between phases in
terms of the phase diagram, Fig. \ref{fig:phase_diagram_2}, in
variables $T_d$ and $T$, which describe, respectively, the
ensemble of sequences and the ensemble of conformations for any
given sequence.  The relevant phases in the diagram are named,
respectively, liquid-like globule, glassy globule, and folded
globule.  We remind to the reader that liquid-like globule is the
state where great many conformations contribute to the partition
function; glassy globule is dominated by one or a few
conformations, but those unrelated to the target conformation
$\star$; and folded globule is dominated by the target
conformation $\star$.  It is fairly obvious and illustrated in
Fig. \ref{fig:phase_diagram_2}, that surface solvation effects
do not change the topology of the phase diagram, but does affect
the specific positions and shape of the corresponding phase
transition lines; these surface-driven changes are the subject of
our interest in this section.

The temperatures of the transitions from liquid-like to glass-like
and to folded globules are called glass temperature $T_{g}$ and
folding temperature $T_{f}$, respectively. Our goal is to analyze
the role of surface solvation effects and design on both $T_{g}$
and $T_{f}$. In other words, we want to calculate how surface
corrections to $T_f$ and $T_g$ depend on the design temperature
$T_d$.

As regards the third phase transition line, that between folded
and glassy globule phases, this line must be vertical in phase
diagram.  Indeed, both folded and glassy globules are zero entropy
states, the transition between them cannot be driven by
temperature change. On the phase diagram, like in Fig. \ref{fig:phase_diagram_2}, 
the corresponding phase boundary must
be represented by the line parallel to the temperature axis.  This
argument does not rely on the volume approximation, and,
therefore, remains valid independently of the surface solvation
effects.  Therefore, this line of phase transition is entirely
described by the value of design temperature at which there is the
triple point, we call it $T_d^{(3)}$.  We want to calculate also
the surface contribution to this quantity.

\begin{figure}
\centering{\scalebox{0.5}{\includegraphics{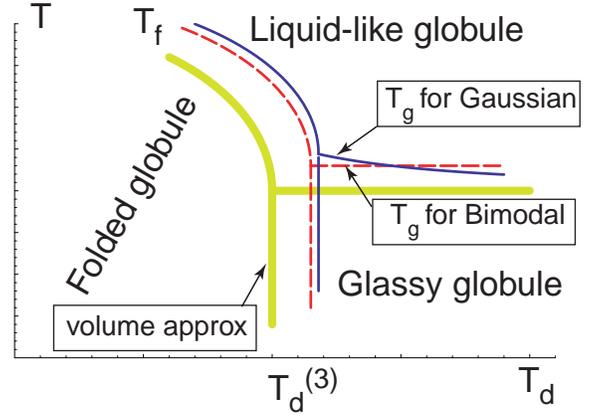}}}
\caption{(Color online) Phase diagram of the heteropolymer system.
There are three phases: random liquid-like globule, frozen glassy
globule and folded globule. Surface term shifts the phase diagram
for volume approximation. For bimodal distribution, glass
temperature is design independent, while in Gaussian distribution,
glass temperature increases for better design condition.}
\label{fig:phase_diagram_2}
\end{figure}

The way we approach the phase diagram is based on the idea that
for any frozen globule phase, whether glassy or folded, the free
energy coincides with energy, because entropy vanishes given the
number of contributing states of order unity.  On the other hand,
the free energy of the liquid-like globule we can find due to the
property of REM that quenched averaged free energy is equal to the
annealed average above the glass temperature.  Therefore, what
we shall do is to compute the annealed average free energy and to
find temperature of entropy ``catastrophe'' - at which entropy
vanishes; that is the glass temperature.  Of course, our goal
is to address surface terms and design terms in this procedure.
Following this program, we write the annealed average as a
functional of the surface monomer distribution $f(\sigma)$ as
\begin{eqnarray}
F \{f\} &=& - T \ln \sum_{{\rm conf}} e^{-E/T} \nonumber\\
&\simeq& - T \ln \int e^{ sN - K\omega_G + N s \{ f(\sigma) \} }
w_G(E) e^{-E/T} dE \nonumber\\
& \simeq & \overline F + F_{\rm surf} \{ f \} \ ,
\end{eqnarray}
where
\begin{equation}
\frac{\overline F} {N} = \overline B - sT - \frac{\delta \! B^2} {2T}
\label{eq:bulk_free_energy}
\end{equation}
is the bulk contribution to the free energy, and
\begin{eqnarray}
\frac{F_{\rm surf} \{ f \}} {K}
&=& \overline\omega T - \frac{\delta \! B^2} {T} \int \sigma^2 p^{\star} (\sigma) d\sigma
+ \frac{N} {K} T \int d\sigma p_{\rm tot}(\sigma) \nonumber\\
& & \times \left[ \phi\left( \eta(\sigma) - \ln \frac{1-\phi} {\phi} \right)
+ \ln (1-\phi) \right]
\label{eq:surface_free_energy}
\end{eqnarray}
is the surface contribution to free energy. In Eq. (\ref{eq:surface_free_energy}),
\begin{eqnarray}
\eta(\sigma) &=& \frac{\delta \! B^2} {T^2} \sigma^2 - \frac{\Gamma} {T} \sigma \ , \nonumber\\
p^{\star} \left( \sigma \right) &=& \frac{ p \left( \sigma \right) \exp \left( \frac{ \Gamma^{\prime d} }
{T_d} \sigma \right) }
{\int p \left( \sigma \right) \exp \left( \frac{ \Gamma^{\prime d} } {T_d} \sigma \right) d\sigma} \ .
\end{eqnarray}
In deriving the free energy, we have used the saddle point approximation since above glass temperature,
free energy is dominated by the saddle point of the partition function.

Optimizing $F_{\rm surf} \{f\}$ yields
\begin{equation}
\phi^{\star} = \frac{1} {1 + \Lambda e^{\eta(\sigma)}} = \frac{K
f^{\star}(\sigma)} {N p_{\rm tot}(\sigma)} \ ,
\end{equation}
and then optimal value of $F_{\rm surf} \{f\}$ evaluates to
\begin{eqnarray}
& & \frac{ F_{\rm surf} \{ f^{\star} \} } {K T} = 1 + \frac{\int
p_{\rm tot} (\sigma) \ln \frac{\Lambda e^{\eta(\sigma)}} {1 +
\Lambda e^{\eta(\sigma)}} d\sigma } {\int
\frac{p_{\rm tot}(\sigma)} {1 + \Lambda e^{\eta(\sigma)}} d\sigma} \nonumber\\
& & -   \frac{\delta \! B^2} {T^2} \int \sigma^2 p^{\star} (\sigma)
d\sigma - \ln \int \frac{\Lambda p_{\rm tot} (\sigma)} {1 + \Lambda
e^{\eta(\sigma)}} d\sigma     \ .
\label{eq:surface_free_energy_optimized}
\end{eqnarray}
These results are similar to those obtained in the work
\cite{Solvation_random}, except we have now non-random sequences,
as manifested by the dependence on $T_d$. In that work
\cite{Solvation_random}, solvation effect in random sequences was
treated as the response of the globule to the surface solvation
``field''.  The linear response regime is characterized by the
statistical independence of disordered parts of surface and volume
energies.  Saturation regime is characterized by the depletion
effect, when preferential solvation of a certain monomer species
exhausts these monomers from the globule.  For completeness, there
is also a narrow range of the so-called weak response regime.
Neither weak response nor saturation regimes are present for the
black-and-white polymer model, with bimodal distribution of
monomer types.

\subsection{No depletion: Glass temperature}

Let us first consider the case when solvation doesn't cause the
depletion of any monomer type, that is, $\phi (\sigma) \ll 1$ at
every $\sigma$, then
\begin{equation}
\frac{F_{\rm surf} \{f^{\star}\}} {K T} \simeq - \frac{\delta \!
B^2} {T^2} \int \sigma^2 p^{\star}(\sigma) d\sigma - \ln \int
p_{\rm tot}(\sigma) e^{-\eta(\sigma)} d\sigma  \ .
\end{equation}
First we will discuss how the glass temperature $T_g$ is affected
by the surface energy, relative to the volume approximated value
$T_{\rm fr} = \delta \! B / \sqrt{2 s}$. Glass temperature must be
determined from by the condition $- \left. \frac{\partial F\{f\}}
{\partial T} \right |_{T_g}=0$. Since $- \left. \frac{\partial
\overline F} {\partial T} \right |_{T_{\rm fr}}=0$, we can write
(denoting temperature derivatives by prime sign) $\Delta T_g \simeq
T_g - T_{\rm fr} \simeq  - \left. \frac{F_{\rm surf}^{\prime}} {
\overline F^{\prime\prime} } \right|_{T_{\rm fr}}$.  Then
\begin{eqnarray}
\frac{\Delta T_g}{T_{\rm fr}} &\simeq& \frac{K}{N} \left[
\frac{\Gamma} {2s T_{\rm fr}} \frac{\int \sigma p(\sigma)
e^{-\eta_{\rm fr}(\sigma)} d\sigma} {\int
p(\sigma) e^{-\eta_{\rm fr}(\sigma)} d\sigma} \right. \nonumber \\
& & \left. + \int \sigma^2 p^{\star}(\sigma) d\sigma - \frac{2\int
\sigma^2 p(\sigma) e^{-\eta_{\rm fr}(\sigma)} d\sigma } 
{\int p(\sigma) e^{-\eta_{\rm fr}(\sigma)} d\sigma} \right. \nonumber \\
& & \left. - \frac{1} {2s} \ln \int p(\sigma) 
e^{-\eta_{\rm fr}(\sigma)} d\sigma \right] \ ,
\label{eq:Delta_T_g}
\end{eqnarray}
where we have replaced $p_{\rm tot}(\sigma)$ with $p(\sigma)$
because $\Delta T_g$ itself is already of order ${\cal O}\left( K
/ N \right)$, and we neglect any higher order corrections. Design
effect is present here through $p^{\star}(\sigma)$.

Relative to no solvation case, the order ${\cal O} (K/N)$ correction
to glass temperature is positive; in other words, glass
temperature is increased due to the solvation effect, so the surface
effect makes ground state more stable.

In the following part, we will further simplify and discuss $\Delta
T_g$ using the examples of $p(\sigma)$.

\subsubsection{Example: bimodal distribution}

For bimodal $p(\sigma)$, $\int \sigma^2 p^{\star}(\sigma)d\sigma =
1$, and design effect doesn't show up in glass temperature,
\begin{eqnarray}
\Delta T_g &=& \frac{K} {2sN} T_{\rm fr} \left[\frac{\Gamma} {T_{\rm fr}} \tanh\left(\frac{\Gamma} {T_{\rm fr}} \right)
- \ln\cosh\left(\frac{\Gamma} {T_{\rm fr}}\right)\right] \nonumber\\
& \simeq & \frac{K}{N}
\begin{cases}
\frac{ \Gamma^2} {2 \sqrt{2 s}  \delta \! B} & \text{when $\frac{\Gamma}{T_{\rm fr}}\ll 1 $}\\
\frac{ \ln 2 } {(2 s)^{3/2} } \delta \! B & \text{when
$\frac{\Gamma}{T_{\rm fr}}\gg 1 $}
\end{cases} \ .
\label{eq:freezing_temperature_bimodal}
\end{eqnarray}
When the solvation strength $\Gamma$ is small, $\Gamma / T_{\rm fr}
\ll 1$, this corresponds to the statistical independent region, in
which surface term and volume term in a heteropolymer globule are
roughly statistically independent. In this region, $\Delta T_g
\propto \Gamma^2$.

The region of $\Gamma / T_{\rm fr} \gg 1$ is saturation region. When
solvation strength $\Gamma$ is so large, essentially all the surface
monomers are of hydrophilic type, as can be clearly seen from
surface monomer distribution Eq. (\ref{eq:surface_bimodal}): when
$\Gamma / T_{\rm fr}\gg 1$, surface monomer distribution function is
dominated by hydrophilic term $f(+1)$.  In this regime, $\Delta T_g$
becomes independent of $\Gamma$, because $\Gamma$ is already so large
that all surface places are occupied by hydrophilic monomers and
further increase of $\Gamma$ cannot change anything.

\subsubsection{Example: Gaussian distribution}

For Gaussian $p(\sigma)$, we can write $\int \sigma^2
p^{\star}(\sigma) d\sigma = \left(\Gamma^{\prime d} / T_d\right)^2 +
1 \simeq \left(\Gamma^d / T_d\right)^2 \left(1 + 2z \delta \! B^d /
T_d \right) + 1 \simeq \left(\Gamma^d / T_d\right)^2  + 1$, where
the asymptotic form comes from the fact that $T_d /(z \delta \! B^d)
\gg 1$ since we  work in the regime of $T_d > T_d^{(3,0)} = \delta
\! B^d / \sqrt{2s}$, where $T_d^{(3,0)}$ is the triple point in
volume approximation.  Then we have
\begin{equation}
\Delta T_g  \simeq  \frac{K}{N} T_{\rm fr} \left[
\left(\frac{\Gamma^d} {T_d}\right)^2 + 6 s  + \frac{ \Gamma^2} {2
\delta \! B^2}\right] \ , \label{eq:freezing_T_Gauss}
\end{equation}
where we also used the fact that $s \ll 1$.  As in the work
\cite{Solvation_random}, system with Gaussian distributed $\sigma$,
unlike bimodal one, has the weak response regime at very small
$\Gamma$, when $\Gamma / T_{\rm fr} \ll s$; in this regime, surface
solvation is insignificant.  The region of $ \Gamma /T_{\rm fr} \gg
\sqrt{24} s$ is the regime where volume and surface disorder are
statistically independent, and the result in this region, in terms
of dependence on $\Gamma$, is similar to that of the bimodal
distribution.

Of course, the major difference from the bimodal example is that in
Gaussian case, there is an additional term due to sequence design.
That term increases glass temperature, which means design, as
usually, makes for a more stable ground state.

\subsection{Triple point}

Now let us consider the folded region, and begin with the
surface-corrected triple point $T_d^{(3)}$; we want to see its
change $\Delta T_d^{(3)} = T_d^{(3)} - T_d^{(3,0)}$ relative to
$T_d^{(3,0)} = \delta \! B^d / \sqrt{2 s}$ determined in volume
approximation.  In general, triple point is determined from the
condition that glass temperature equals folding temperature, $T_f =
T_g $.  Glass temperature, along with its surface corrections, is
already known to us, see formula (\ref{eq:Delta_T_g}) or its
simplified versions (\ref{eq:freezing_temperature_bimodal}) and
(\ref{eq:freezing_T_Gauss}).  The folding temperature $T_f$ should
be calculated from $\langle E_{\star} ({\rm seq}) \rangle = F \{
f^{\star} \}$.  We take averaged ground state energy from formula
(\ref{eq:averaged_ground_state_energy}) and we take $F \{ f^{\star}
\}$ from Eqs. (\ref{eq:bulk_free_energy}) and
(\ref{eq:surface_free_energy_optimized}); in the latter (which is
the surface part) we must neglect all order ${\cal O}(K/N)$
corrections.  The result reads
\begin{eqnarray}
\Delta T_d^{(3)} &\simeq& \frac{K} {N} T_d^{(3,0)}
\left[\frac{\Gamma \Gamma^d} {\delta \! B \delta \! B^d}
- \int \sigma^2 p^{\star}(\sigma) d\sigma \right. \nonumber\\
& & \left. - \frac{1} {2s} \ln \int p(\sigma) e^{-\eta_{\rm fr} (\sigma)}
d\sigma \right] \ .
\label{eq:triple_point_general}
\end{eqnarray}
From this formula, it is not even clear whether $\Delta T_d^{(3)}$
is positive or negative.  As with other cumbersome results, let us
look at the specific examples of $p( \sigma )$.

\subsubsection{Example: bimodal distribution}

For bimodal distribution $p(\sigma)$, we have
\begin{eqnarray}
& & \Delta T_d^{(3)} \simeq \frac{K} {N} T_d^{(3,0)} \left[
\frac{\Gamma \Gamma^d} {\delta \! B \delta \! B^d}
- \frac{1} {2s} \ln \cosh \left( \frac{\Gamma} {T_{\rm fr}} \right) \right]\nonumber\\
&\simeq&
\begin{cases} \frac{K} {N} T_d^{(3,0)} \frac{\Gamma} {\delta \! B} \left( \frac{\Gamma^d} {\delta \! B^d}
- \frac{\Gamma} {2 \delta \! B} \right)  & \text{when $\frac{\Gamma} {T_{\rm fr}} \ll 1$} \\
  \frac{K} {N} T_d^{(3,0)} \frac{\Gamma} {\delta \! B} \left( \frac{\Gamma^d} {\delta \! B^d}
 - \frac{1} {\sqrt{2s}} \right)   & \text{when $\frac{\Gamma} {T_{\rm fr}} \gg 1$}
\end{cases} \ . \label{eq:triple_point_bimodal}
\end{eqnarray}
We see that the design effect increases $\Delta T_d^{(3)}$, pushes
triple point to the right on the phase diagram Fig. \ref{fig:phase_diagram_2}, 
while the solvation effect acts in the
opposite direction.  Interestingly, in the statistical independence
regime, when $\Gamma / T_{\rm fr} \ll 1$, the sign of $\Delta
T_d^{(3)}$ is determined by the competition of the design term
$\Gamma^d / \delta \! B^d$ and folding term $\Gamma / \delta \! B$.
Specifically, large design solvation strength $\Gamma^d / \delta \!
B^d \gg \Gamma / \delta \! B$ would make $\Delta T_d^{(3)}
> 0$ and in this sense the design makes folded state more stable.  
From Eq. (\ref{eq:freezing_temperature_bimodal}), we already know that
glass temperature is independent of $\Gamma$ in saturation region
$\Gamma / T_{\rm fr} \gg 1$. Not surprisingly, in this region, the
sign of $\Delta T_d^{(3)}$ is also independent of $\Gamma$.

\subsubsection{Example: Gaussian distribution}

When $p(\sigma)$ is Gaussian, we get
\begin{eqnarray}
\Delta T_d^{(3)} &\simeq& \frac{K} {N} T_d^{(3,0)} \left[
\frac{\Gamma {\Gamma}^d} {\delta \! B \delta \! B^d}
- \frac{\Gamma^2} {2 \delta \! B^2} - 2s \right. \nonumber\\
& & \left. - 2s \left( \frac{\Gamma^d} {\delta \! B^d} \right)^2 (1
+ 2z\sqrt{2s}) \right] \ . \label{eq:triple_point_Gaussian}
\end{eqnarray}
The interesting and rather unexpected result is that the design
effect, when it is \emph{very} strong, might lead to reduction of
$T_d^{(3)}$; in other words, it might have an adverse effect on the
stability of the folded phase.  Inspection of the origin of the
negative term $ \propto - (\Gamma_d / \delta \! B^d)^2$ shows that
its origin is due to the fact that very strong solvation effect in
design brings in a significant fraction of very strongly solvophilic
monomers; even though only small fraction of them subsequently turns
out inside the globule in the folded state, they nevertheless make
the destabilizing effect on the globule.  We emphasize that such
danger exists only when solvation effect in design is so strong that
not only $\Gamma^d / \delta \! B^d > \Gamma / \delta \! B$, but $s
\Gamma^d / \delta \! B^d > \Gamma / \delta \! B $.  It is unclear if
such situation is realistic.

\subsection{Folding temperature}

Next let us consider the folding temperature $T_f$ away but not far
from the triple point. When $T_d < T_d^{(3)}$, we have
\begin{equation}
\frac{\delta \! B \delta \! B^d + \frac{K} {N} \Gamma \Gamma^d }
{T_d} = \left. \left(s T + \frac{\delta \! B^2} {2 T} + F_{\rm surf}
\right) \right|_{T = T_f} \ .
\end{equation}
In the vicinity of the triple point, when $T_d = T_d^{(3)} - \Delta
T_d$, where $\Delta T_d \ll \frac{K}{N} T_d^{(3)}$, we have $F_{\rm
surf}|_{T=T_f} \simeq F_{\rm surf}|_{T=T_g, T_d=T_d^{(3,0)}}$, or
\begin{equation}
\frac{\delta \! B \delta \! B^d } {T_d^{(3)}} \frac{\Delta T_d} {T_d^{(3)}}
\simeq  s T_f + \frac{\delta \! B^2} {2 T_f} - s T_g - \frac{\delta \! B^2} {2 T_g} \ ,
\end{equation}
which yields upon some algebra
\begin{equation}
T_f \simeq T_g \left[1 + \sqrt{\frac{ \delta \! B \delta \! B^d} {s
T_d^{(3)} T_g} \left(1 - \frac{T_d} {T_d^{(3)}} \right)} \right] \ .
\end{equation}

In Fig. \ref{fig:phase_diagram_2}, the phase diagram of the system
is sketched. In the regime of temperature $T$ below $T_f$, the
designed sequences will thermodynamically stable when folded to the
target state. In the regime of $T_d > T_d^{(3)}$, sequences obtained
will be either in frozen glassy state or in random liquid-like
globule.

\subsection{Depletion effect}

In preceding sections, we assumed no depletion effect. Here we will
see what happens if there is depletion. Depletion of monomers may
happen for Gaussian $p(\sigma)$, while in bimodal case, there is no
depletion since the number of monomers for each monomer type is
abundant. When depletion occurs, $\phi = 1$ for $\sigma \geq
\sigma_m$, and $\phi = 0$ for $\sigma < \sigma_m$, and this leads to
\begin{equation}
\int_{\sigma_m}^{\infty} p_{\rm tot} (\sigma) d\sigma= K/N \ .
\end{equation}
The integration result is
\begin{equation}
\underbrace{ {\rm erfc} \left( \frac{\sigma_m - \frac{
\Gamma^{\prime d}}{T_d}} {\sqrt{2}} \right) - {\rm erfc}
\left(\frac{\sigma_m} {\sqrt{2}} \right) }_{ >0 } - 2 = - \frac{N}{K}
{\rm erfc} \left( \frac{\sigma_m} {\sqrt{2}} \right) \ .
\end{equation}

For the case of no design, $\Gamma^{\prime d} = 0 $,
\begin{equation}
{\rm erfc} \left(\frac{\sigma_m^0} {\sqrt{2}} \right) = \frac{2K}
{N} \ .
\end{equation}
Therefore, we have ${\rm erfc} \left(\sigma_m / \sqrt{2} \right) <
{\rm erfc} \left( \sigma_m^0 /\sqrt{2} \right)$, and this gives
$\sigma_m > \sigma_m^0$, so with design, the surface monomers are
more hydrophilic than without design. This makes physical sense and
this is also consistent with no depletion case $\phi \ll 1$, in
which design favors the hydrophilic monomers on the surface.

\section{Sequence space entropy}\label{sec:sequence_entropy}

Sequence design, when it is realized computationally, or if it could
be realized experimentally, helps finding sequences with
particularly low ground state energy.  But of course there is a
limit - there is always a sequence whose ground state energy is the
lowest among all sequences, and, therefore, no design can possibly
produce any sequence with lower energy. More generally and more
practically, the lower ground state energy we want to obtain, the
fewer sequences exist which can meet our demand.  One may want to
know how many sequences are there to choose from with any given
ground state energy.  Design paradigm provides the general method to
solve such problem.  Indeed, we can compute the sequence space
entropy (which is just the logarithm of the number of relevant
sequences) as a function of $T_d$ and as we also know the average
ground state energy as a function of $T_d$, we can determine the
number of sequences as depends on their ground state energy.  This
procedure in volume approximation is described in the work
\cite{RMP}.  Here we want to consider how it is affected by the
surface solvation effect.

In principle, sequence space entropy depends quite strongly on the
target state fold here denoted as $\star$, this dependence is called
designability of the fold (see, for example, recent work on this
subject \cite{Designability}).  Here, we will neglect this fact.
This is not because designability is unimportant - it is very
important indeed, but our goal is to look at the role of sequence
solvation effect, so to make this task manageable, we have to
sacrifice the designability issue as a zeroth approximation.

To find sequence space entropy, we consider sequence space free
energy $ -T_d\ln Z$, where $Z = \sum_{{\rm seq}} \exp\left[ -
H^d({\rm seq}, {\star}) / T_d \right]$. Note that design is to a
certain conformation $\star$, so in the partition function here, the
summation runs over sequences. Sequence space entropy per monomer
$s_{\rm seq} $ can then be found using high $T_d$ expansion just in
the same way as the calculation of $\langle E_{\star}(seq) \rangle$,
formula (\ref{eq:averaged_ground_state_energy}). The result is given
by
\begin{equation}
s_{\rm seq} = -\frac{\partial \left( -T_d \ln Z \right)} {\partial
T_d} \simeq  \ln q - \frac{ \delta \! {B^d}^2 + \frac{K} {N}
{\Gamma^d}^2} {2T_d^2} \ , \label{eq:sequence_entropy}
\end{equation}
where $q$ is the effective number of `letters in the alphabet'
determined from the total number of possible sequences of length
$N$: ${\cal N}_{\rm seq} = q^N$.  It is not difficult to show that
$q = - \sum_{\sigma} p(\sigma) \ln p(\sigma)$ (see Eq.
(\ref{eq:propensities}); $q \approx 18$ for real proteins).

The number of sequences is maximal if we impose no constraints on
the quality of design, which means sequence entropy has to be
maximal when $T_d$ is at the triple point.  Therefore, we can
compute $s_{{\rm seq}}^{\rm max}$ using formula
(\ref{eq:sequence_entropy}) at $T_d = T_d^{(3)}$. The result reads:
\begin{equation}
s_{{\rm seq}}^{\rm max} \simeq \ln q - s \left[ 1 - \frac{2 \Delta
T_d^{(3)}} {T_d^{(3,0)}} + \frac{K {\Gamma^d}^2} {N \delta {\!
B^d}^2} \right]  \ , \label{eq:seq_max}
\end{equation}
where the ratio $\Delta T_d^{(3)} / T_d^{(3,0)}$ should be taken
from Eq. (\ref{eq:triple_point_general}) or from the simplified
versions of it (\ref{eq:triple_point_bimodal}) or
(\ref{eq:triple_point_Gaussian}).

First, in the volume approximation, when there is no surface term,
we have $s_{{\rm seq}}^{\rm max} = \ln q - s$. This result is a very
natural consequence of our neglect of the difference in
designabilities between different folds. Indeed, volume
approximation of $s_{{\rm seq}}^{\rm max}$ indicates that the number
of sequences that can be designed for a given conformation $\star$
is $e^{Ns_{{\rm seq}}^{\rm max}} = {\cal N}_{{\rm seq}} / {\cal
N}_{{\rm conf}}$, which means that all ${\cal N}_{{\rm seq}} =
q^{N}$ sequences are equally distributed between ${\cal N}_{\rm
conf} = e^{s N}$ conformations.  This is because the fraction of
sequences with ground state energy above $\langle E_{\star}(seq)
\rangle$ (\ref{eq:averaged_ground_state_energy}) is extremely small,
see appendix \ref{sec:ground_state}, so practically all sequences
have their ground state energy around $\langle E_{\star}(seq)
\rangle$.

Second, we look at the role of surface effect. For simplicity, we
restrict consideration to the most typical regime of statistical
independence between surface and bulk contributions. For both
bimodal distribution and Gaussian distribution, plugging $\Delta
T_d^{(3)}$ into Eq. (\ref{eq:seq_max}), we get the following simple
result:
\begin{equation}
s_{\rm seq}^{\rm max} = \ln q - s - \frac{s K} {N} \left[
\frac{\Gamma} {\delta \! B} - \frac{\Gamma^d} {\delta \! B^d}
\right]^2  \ .
\end{equation}
This tells us that the surface solvation effect reduces the number
of sequences, $s_{\rm seq}^{\rm max} < \ln q -s$.  This happens
because some of the sequences, with inadequate supply of hydrophilic
monomers, have their ground state energies above $\langle
E_{\star}(seq) \rangle$ when we look at them carefully enough to
notice their surface energy.  Accordingly, the fraction of sequences
with ground state at $\star$ is below its `fair' share of $e^{(\ln
q-s)N}$ and even maximal sequence space entropy falls short of its
volume approximated value $\ln q - s$.  Only very careful design, at
which $\Gamma^d/\delta \! B^d = \Gamma/\delta \! B$, would be able
to provide the ensemble of sequences adequate to their solvation
conditions, in which case the solvation effect does not increase
energy and does not rule any sequences out of the competition.
Notice that the condition $\Gamma^d/\delta \! B^d = \Gamma/\delta \!
B$ does not involve design temperature, it specifies only the
balance of solvation and bulk heteropolymeric effects.

Let us now look at the situation differently, namely, let us write
down the folding temperature $T_f$ in terms of $s_{\rm seq}$ instead
of $T_d$.  Indeed, $T_d$ is a purely technical concept which may or
may not directly correspond to the experimental reality; for
instance, design can be controlled by some analog of solvent quality
instead of temperature.  At the same time, sequence entropy is a
very clear quantity, it is the number of sequences whose ground
state stability corresponds to the temperature $T_f$.  Simple
algebra shows that
\begin{equation}
T_f  =  T_g \left[ 1 + \sqrt{\frac{\delta \! B \delta \! B^d} {s T_g
T_d^{(3)}} \left(1 - \sqrt{\frac{ \ln q - s_{\rm seq}^{\rm max} } {
\ln q - s_{\rm seq} }} \right) } \right] \ .
\end{equation}
This allows to re-interpret phase diagram, Fig.
\ref{fig:phase_diagram_2}, with sequence entropy on the horizontal
axis.

Finally, it is known \cite{Shakhnovich_review_2006} that the quality
of design is best characterized by the energy gap between the energy
of the sequence in its purported target state and the average ground
state energy
\begin{equation} \Delta \epsilon  = \frac{ \left. F\{ f^\star \} \right |_{T_g} - \langle E_{\star}
({\rm seq}) \rangle }{N} \ . \end{equation}
Therefore, we should look at the relation between sequence entropy
and $\Delta \epsilon$.  From the above results, we have found
\begin{equation}
s_{\rm seq}   =  \ln q - s \left( 1 + \frac{\Delta \epsilon}{\sqrt{2
s} \delta \! B} + \frac{K}{N} \zeta \right)^2 \left( 1 + \frac{K}{N}
\xi \right) \ ,
\end{equation}
where the solvation related coefficients are given by
\begin{equation} \xi =  \frac{ {\Gamma^d}^2} { {\delta \! B^d}^2} - 2\frac{ \Gamma
\Gamma^d}{ \delta \! B \delta \! B^d}  \end{equation} and
\begin{equation} 
\zeta  = \int \sigma^2 p^\star (\sigma) d\sigma  
+ \frac{1}{2 s} \ln \int p(\sigma) e^{-\eta_{\rm fr} (\sigma)} d\sigma \ .
\end{equation}
There are less sequences available for larger energy gap design.

\subsubsection{Example: bimodal distribution}

When $p(\sigma)$ is bimodal,
\begin{eqnarray}
\zeta =  \frac{1}{2 s} \ln \cosh \frac{\Gamma} {T_{\rm fr}}
\ .
\end{eqnarray}

\subsubsection{Example: Gaussian distribution}

When $p(\sigma)$ is Gaussian,
\begin{eqnarray}
\zeta = \frac{\Gamma^2} {2\delta \! B^2} + 2s 
+ 2s \left(\frac{\Gamma^d}{\delta \! B^d} \right)^2  \ .
\end{eqnarray}

In either case, we see that the number of available sequences drops
dramatically as we increase their desired quality by choosing a
larger $\Delta \epsilon$.

\section{Conclusion}

In this paper, we examined the interplay of surface solvation
effects and sequence design for protein-like heteropolymer globule.
Ideologically, our treatment of disordered sequences  followed the
theoretical studies of heteropolymer folding in the works
\cite{BW,REM_Derrida,BJ,SG_IndependentInt}, and in our treatment of
preferential solvation we used the approach of the work
\cite{Solvation_random}.  What we did is we applied REM in the new
challenging context.

As in the volume approximation, designed sequences in the target
conformation have lower energy than random sequences. This is not
surprising: this is after all the sole purpose of design. Less
obvious, we found that the role of preferential solvation for the
design itself might be controversial.  The problem is that when
design conditions favor too strongly the hydrophilicity of the
surface monomers, these monomers can have an adverse effect on the
overall composition of the sequence and then disrupt the favorable
arrangement of contacts inside the globule.

Speaking about phase diagram of the heteropolymer globule, we found
that surface solvation effect operates differently for the two most
typical examples of monomer composition.  If there are only two
types of monomers, then glass transition temperature remains
independent of the design condition, as it was found in volume
approximation.  But this is not longer the case when there is a wide
Gaussian distribution of monomer types; in this case, design brings
in a noticeable fraction of very hydrophilic monomers from the tail
of hydrophilicity distribution, and they do affect the glass
transition.

To conclude, our study shows it possible to incorporate preferential
solvation effects into the the REM-based heteropolymer theory, and
some of the obtained results are quite delicate and unexpected.  In
reality, the role of surface in molecules of realistic sizes is
quite significant, so the effects which were examined here on
perturbative level, considering surface contributions ${\cal
O}(K/N)$ very small, might be quite substantial and very important.

\acknowledgements The work of both authors was supported in part by
the MRSEC Program of the National Science Foundation under Award
Number DMR-0212302.

\appendix

\section{Probability distributions of the low energy states in REM}\label{sec:REM_distrib}

To make this work self-contained, we review here the probability
distributions of the low lying states in the REM.  Further details
on this subject can be found in \cite{Bouchaud_Mezard}.

\subsection{Ground state energy}\label{sec:ground_state}

Consider some ${\cal M}$ (${\cal M}=e^{sN}$) statistically
independent energy levels, each distributed with probability
density $w(E)$. The question is this:  what is the probability
distribution, ${\cal W}(E)$ of the lowest among these ${\cal M}$
levels? The general answer, due to the statistical independence,
reads
\begin{equation} {\cal W} (E) = {\cal M} w(E) \left( \int_E^{\infty}
w(E^{\prime}) dE^{\prime} \right)^{{\cal M}-1} \ . \label{eq:exact}
\end{equation}
Here, $w(E)$ is the probability that there is a level at $E$,
$\left( \int_E^{\infty} w(E^{\prime}) dE^{\prime} \right)^{{\cal
M}-1}$ is the probability that all other ${\cal M}-1$ levels
happen to be above $E$, and factor ${\cal M}$ reflects the fact
that any of the ${\cal M}$ levels may play the role of the lowest
one.

With a large number ${\cal M} \gg 1$ of levels taken from the
distribution $w(E)$, the lowest one of them will surely be located
somewhere in the low energy tail of the distribution $w(E)$.  In
this region, $\int_{E}^{\infty} w(E^{\prime}) d E^{\prime}$ is
very close to unity, or in other words $\int_{E}^{\infty}
w(E^{\prime}) d E^{\prime} = 1 - \int_{-\infty}^{E} w(E^{\prime})
d E^{\prime} \simeq \exp \left[ - \int_{-\infty}^{E} w(E^{\prime})
d E^{\prime} \right] $. Neglecting also $1$ compared at ${\cal M}$
in ${\cal M}-1$, we then re-write eq. (\ref{eq:exact}):
\begin{equation} {\cal W} (E) \simeq {\cal M} w(E) e^{-{\cal M}
\int_{-\infty}^E w(E^{\prime}) dE^{\prime} } \ .
\label{eq:almost_exact} \end{equation}

Where is the maximum of ${\cal W}(E)$, what is the most probable
ground state energy, $E_m$?  The condition ${\cal
W}^{\prime}(E)=0$ yields
\begin{equation} {\cal M} w(E_m)  = \left( \ln w(E_m) \right)^{\prime}
\label{eq:Em} \ . \end{equation}
Apart from the logarithmic corrections, this returns the familiar
condition traditionally written in a careless form ${\cal M} w(E)
\simeq 1$ in which units do not match.

We now remember that $w(E)$ is Gaussian,
\begin{equation} w(E) = \frac{1}{\sqrt{ 2 \pi N \delta \! B^2}} e^{-E^2/2 N
\delta \! B^2 } \ .  \end{equation}
In this case formula (\ref{eq:Em}) reads
\begin{equation} {\cal M} w(E_m)  = - E_m / N \delta \! B^2 \ , \end{equation}
which then implies
\begin{equation} E_m  =  - \sqrt{2 N \delta \! B^2 \left\{  \underbrace{\ln
{\cal M}}_{Ns} - \ln \left[ \frac{-E_m}{ N \delta \! B^2} \sqrt{ 2
\pi N \delta \! B^2} \right] \right\}} \end{equation}
and by simple iteration we finally have
\begin{eqnarray}  E_m & \simeq & - \sqrt{2 s} N \delta \! B + \frac{\delta \!
B}{2 \sqrt{2 s}} \ln \left( 4 \pi N s \right) \nonumber \\
& \simeq & - \sqrt{2 s} N \delta \! B + {\cal O}( \ln N )  \ .
\label{eq:EmSAMO} \end{eqnarray}
The leading term here corresponds to dropping the pre-exponential
factor of $w(E)$ when writing ${\cal M} w(E) \simeq 1$, in which
case, of course, units match and the result is correct.

It is fairly obvious, and will be verified instantly, that ${\cal
W}(E)$ is concentrated around $E_m$.  Accordingly, let us assume
$E = E_m + \epsilon$, with $\left| \epsilon \right| \ll \left| E_m
\right|$.  In this small $\epsilon$ range, we can simplify the
Gaussian $w(E)$:
\begin{eqnarray}  {\cal M} w(E) & = & \frac{{\cal M}}{\sqrt{2 \pi N \delta \!
B^2}} e^{-\left(E_m + \epsilon \right)^2 / 2 N \delta \! B^2}
 \nonumber \\ & \simeq & \underbrace{{\cal M}
w(E_m)}_{\sqrt{2s}/\delta \! B} e^{\epsilon E_m/ N \delta \! B^2}
\simeq \frac{1}{T_{\rm fr}} e^{\epsilon /T_{\rm fr} }\ ,
\label{eq:w_at_porog} \end{eqnarray}
where $T_{\rm fr} = \frac{\delta \! B}{\sqrt{2s}}$.  We further
use the Gaussian shape of $w(E)$ and the proper asymptotics of the
error function to write
\begin{equation} {\cal M} \int_{-\infty}^{E} w(E^{\prime}) d E^{\prime} \simeq
{\cal M} w(E) \frac{N \delta \! B^2}{-E} \simeq e^{\epsilon /T_{\rm
fr} }\ . \label{eq:erf_at_porog} \end{equation}
Plugging both (\ref{eq:w_at_porog}) and (\ref{eq:erf_at_porog})
into eq. (\ref{eq:almost_exact}), we arrive at the following
probability distribution of the ground state energy
\begin{equation} {\cal W}(E) \simeq (1/ T_{\rm fr} ) e^{(\epsilon/T_{\rm fr}) -
\exp (\epsilon / T_{\rm fr}) }  \ , \label{eq:WofE} \end{equation}
where, once again,
\begin{equation} \epsilon = (E - E_m) \ , \ T_{\rm fr} = \frac{ \delta \! B}{
\sqrt{2 s} }  \ . \end{equation}
As expected, ${\cal W}(E)$ is not symmetric, it decays much faster
to the right (higher energies) than to the left (lower energies).
However, the characteristic scale in both cases is set by the
quantity $T_{\rm fr} = \delta \! B / \sqrt{2 s}$ and does not
contain $N$ in any form.  It is only the parameter-free ``shape''
$e^{\xi - e^{\xi}}$ that is asymmetric.  It is also satisfying
that ${\cal W}(E)$ Eq. (\ref{eq:WofE}) is correctly normalized.

\subsection{Lowest and second lowest energy states: joint distribution}

Consider now the joint probability distribution of the two lowest
energies.  Exact formula for this joint distribution, similar to
Eq. (\ref{eq:exact}) reads
\begin{eqnarray}  {\cal W}(E_1,E_2) & = & {\cal M} \left( {\cal M} -1 \right)
w(E_1) w(E_2) \nonumber \\ & \times & \left( 
\int_{E_2}^{\infty} w(E^{\prime}) d E^{\prime} \right)^{{\cal M} -2}
\ . \end{eqnarray}
Similar to Eq. (\ref{eq:exact}), we rely here on statistical
independence of energy levels in REM, which implies that we should
consider the independent factors of one energy level present at
$E_1$ (which gives $w(E_1)$), another energy level present at
$E_2$ (yielding $w(E_2)$), times the probability that all other
${\cal M} -2$ energy levels are above $E_2$, and times the
combinatorial factor ${\cal M} \left( {\cal M} -1 \right)$ which
reflects the idea that any two of the ${\cal M}$ levels can play
the role of the lowest and second lowest.

The first step to simplify this distribution is to notice that both
lowest $E_1$ and second lowest $E_2$ energy levels are practically
always located far in the lower energy tail of the distribution
$w(E)$.  Similar to Eq. (\ref{eq:almost_exact}), we then write
\begin{eqnarray}  {\cal W}(E_1,E_2) & = & {\cal M}^{2} w(E_1) w(E_2) \nonumber \\
& \times & \exp \left[ - {\cal M} \int_{-\infty}^{E_2} w(E^{\prime})
d E^{\prime} \right] \ . \end{eqnarray}
The next step is to rely on equations (\ref{eq:w_at_porog}) and
(\ref{eq:erf_at_porog}).  Denoting
\begin{equation} E_1 = E_m + \epsilon_1 \ , \ \ {\rm and} \ \ \ E_2 = E_m +
\epsilon_2 \ , \end{equation}
and assuming $\left| \epsilon_{1,2} \right| \ll \left| E_m
\right|$, we arrive at
\begin{equation} {\cal W}(E_1,E_2) \simeq \frac{1}{T_{\rm fr}^2}
e^{(\epsilon_1/T_{\rm fr}) + (\epsilon_2/T_{\rm fr}) - \exp
(\epsilon_2 /T_{\rm fr} )} \ . \end{equation}
\subsection{Some derivative probability distributions}

It is a wonderful rewarding exercise to establish that integrating
out the first excited state $E_2$ returns the ground state
distribution (\ref{eq:WofE}), as it must: $\int_{E_1}^{\infty}
{\cal W}(E_1,E_2) dE_2 = {\cal W} (E_1)$.

\subsubsection{Probability distribution of the second lowest
state}

For the second lowest energy, we have by simple integration
\begin{eqnarray}  {\cal W}_2 (E) & = & \int_{-\infty}^{E} {\cal W}(E_1,E)
dE_1  \nonumber \\ & = & \frac{1}{T_{\rm fr}} e^{2(\epsilon
/T_{\rm fr}) - \exp ( \epsilon/T_{\rm fr}) } \ .\eea
This distribution is different from that of the lowest energy
state, Eq. (\ref{eq:WofE}), as it is clear from the Figure
\ref{fig:Dva_Raspred}.  It is hardly a surprise that the second
lowest energy distribution is somewhat shifted towards larger
energies compared at the distribution of the ground state.

\begin{figure}
\centerline{\scalebox{0.45} {\includegraphics{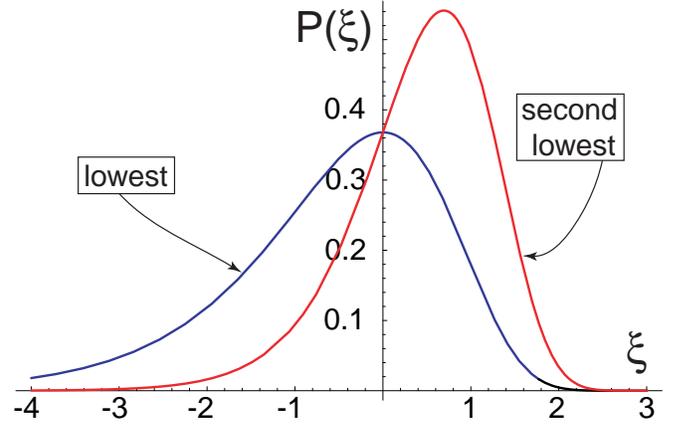}}}
\caption{Probability distributions for the lowest and second
lowest energy states in REM.  Horizontal axes is $\xi = \epsilon /
T_{\rm fr}$, and vertical axes is the probability density for this
quantity $\xi$ (so the area under each curve in the figure is
unity).} \label{fig:Dva_Raspred}
\end{figure}

\subsubsection{Conditional probability of $E_2$ at given $E_1$}

Suppose ground state energy is fixed at $E_1$, what is the
probability distribution of $E_2$ at the given $E_1$?  According
to the general rules of probabilities, this quantity is equal to
\begin{eqnarray}  {\cal W} ( \left. E_2 \right| E_1 ) & =  & \frac{{\cal
W}(E_1,E_2)}{{\cal W}(E_1)} \nonumber \\ & = & \frac{1}{T_{\rm fr}}
e^{(\epsilon_2 / T_{\rm fr}) - \exp (\epsilon_2 /T_{\rm fr}) + \exp
(\epsilon_1 / T_{\rm fr})} \ . \end{eqnarray}
For understanding, it is useful to mention that this quantity is
normalized by the condition $\int_{E_1}^{\infty} {\cal W} ( \left.
E_2 \right| E_1 )  dE_2 =1$.

\subsubsection{Probability distribution for the gap between lowest
and second lowest states}

As regards the gap between two lowest energy states, $\Delta E =
E_2 - E_1$, its probability distribution is also easily found:
\begin{eqnarray}  && {\cal W} (\Delta E) \nonumber \\ & = &
\int_{-\infty}^{\infty} dE_1 \int_{E_1}^{\infty} dE_2 {\cal
W}(E_1,E_2) \delta \left(E_2 - E_1 - \Delta E \right)
\nonumber
\\ & = & \int_{-\infty}^{\infty} {\cal W}(E_2 - \Delta E,E_2) d E_2
\nonumber \\ & \simeq & \frac{1}{T_{\rm fr}^2}
\int_{-\infty}^{\infty} \exp \left[\frac{\epsilon_2 - \Delta
E}{T_{\rm fr}} + \frac{\epsilon_2}{T_{\rm fr}} -
e^{\epsilon_2/T_{\rm fr}}\right] d \epsilon_2 \nonumber \\ &=&
\frac{1}{T_{\rm fr}} e^{-\Delta E/T_{\rm fr}} \int_{0}^{\infty}
\xi e^{-\xi} d \xi = \frac{1}{T_{\rm fr}} e^{-\Delta E/T_{\rm fr}}
\ .\eea
Of course, this is the result for $\Delta E >0$, so in general
probability distribution for the gap reads
\begin{equation} {\cal W}(\Delta E) =
\left\{\begin{array}{lcr} 0& {\rm for} & \Delta E < 0  \\
(1/T_{\rm fr}) e^{-\Delta E/T_{\rm fr}} & {\rm for} & \Delta E > 0
\end{array} \right.
\label{eq:gap_distrib}\ee
It might seem surprising at the first glance that this probability
decays only as exponential, not the sharper $e^{-\exp \xi}$
function.  Indeed, if we imagine gap forming as fixing lowest
energy at the typical place and then looking at the second lowest
level, then the probability of the gap decays in a very sharp,
double exponential way.  However, if we think differently, that
$E_2$ is fixed in a typical place, and then gap is determined by
where is the $E_1$, then the probability of $E_1$ going down is
just exponential.  Thus, from this argument we gain the following
insight.  The probability of the gap is exponential because the
dominant possibility for the gap to be large is having unusually
low $E_1$ rather than high $E_2$.  Thus, large gap is typically
the result of the low ground state.

\subsection{Averages}

Using formula (\ref{eq:WofE}), it is easy to compute the
expectation value of the ground state energy:
\begin{eqnarray}  \left< E_g \right> & = & \int_{-\infty}^{\infty} E {\cal
W}(E) d E \nonumber \\ & = & E_m + T_{\rm fr}
\int_{-\infty}^{\infty} \xi e^{\xi - \exp \xi } d \xi \nonumber
\\ & = & E_m + \frac{\delta \! B}{\sqrt{2 s}}
\underbrace{ \int_{0}^{\infty} \ln \eta e^{-\eta} d \eta
}_{\Gamma^{\prime}(1)\approx - 0.6}
 \ , \label{eq:ground_state} \end{eqnarray}
where $\xi = \epsilon / T_{\rm fr}$ and $\eta = e^{\xi}$.  As a
fact of mathematical curiosity, there appears $\Gamma^{\prime}$,
the derivative of the Euler $\Gamma$-function - rather infrequent
guest in physics calculations.  However exotic mathematically,
this term is smaller than the logarithmic correction term in
(\ref{eq:EmSAMO}) and so it must be neglected if the first
approximation is used for $E_m$.

We can also compute the average value of the second lowest energy:
\begin{eqnarray}  \left< E_2 \right> & = & \int_{-\infty}^{\infty} E {\cal W}_2
(E) d E \nonumber \\ & = & E_m + T_{\rm fr}
\int_{-\infty}^{\infty} \xi e^{2 \xi - \exp \xi } d \xi
\nonumber
\\ & = & E_m + \frac{\delta \! B}{\sqrt{2 s}}
\underbrace{ \int_{0}^{\infty} \eta \ln \eta e^{-\eta} d \eta
}_{\Gamma^{\prime}(2)= 1+ \Gamma^{\prime}(1) \approx  0.4}
 \ . \label{eq:excited_state} \end{eqnarray}
This is greater than the average ground state energy
(\ref{eq:ground_state}) by just the amount $T_{\rm fr} = \delta \!
B / \sqrt{2 s}$.

The latter result is also consistent with the fact that the
average gap, according to eq. (\ref{eq:gap_distrib}) is equal to
\begin{equation} \langle \Delta E \rangle = T_{\rm fr} = \delta \! B / \sqrt{2
s} \ . \end{equation}

\bibliography{solvation}

\begin{thebibliography}{20}
\expandafter\ifx\csname natexlab\endcsname\relax\def\natexlab#1{#1}\fi
\expandafter\ifx\csname bibnamefont\endcsname\relax
  \def\bibnamefont#1{#1}\fi
\expandafter\ifx\csname bibfnamefont\endcsname\relax
  \def\bibfnamefont#1{#1}\fi
\expandafter\ifx\csname citenamefont\endcsname\relax
  \def\citenamefont#1{#1}\fi
\expandafter\ifx\csname url\endcsname\relax
  \def\url#1{\texttt{#1}}\fi
\expandafter\ifx\csname urlprefix\endcsname\relax\def\urlprefix{URL }\fi
\providecommand{\bibinfo}[2]{#2}
\providecommand{\eprint}[2][]{\url{#2}}

\bibitem[{\citenamefont{Bresler and Talmud}(1944{\natexlab{a}})}]{Bresler1}
\bibinfo{author}{\bibfnamefont{S.}~\bibnamefont{Bresler}} \bibnamefont{and}
  \bibinfo{author}{\bibfnamefont{D.}~\bibnamefont{Talmud}},
  \bibinfo{journal}{Doklady AN SSSR} \textbf{\bibinfo{volume}{47}},
  \bibinfo{pages}{326} (\bibinfo{year}{1944}{\natexlab{a}}).

\bibitem[{\citenamefont{Bresler and Talmud}(1944{\natexlab{b}})}]{Bresler2}
\bibinfo{author}{\bibfnamefont{S.}~\bibnamefont{Bresler}} \bibnamefont{and}
  \bibinfo{author}{\bibfnamefont{D.}~\bibnamefont{Talmud}},
  \bibinfo{journal}{Doklady AN SSSR} \textbf{\bibinfo{volume}{47}},
  \bibinfo{pages}{367} (\bibinfo{year}{1944}{\natexlab{b}}).

\bibitem[{\citenamefont{Finkelstein and Ptitsyn}(2002)}]{Ptitsyn_Finkelstein}
\bibinfo{author}{\bibfnamefont{A.~V.} \bibnamefont{Finkelstein}}
  \bibnamefont{and} \bibinfo{author}{\bibfnamefont{O.~B.}
  \bibnamefont{Ptitsyn}}, \emph{\bibinfo{title}{Protein Physics: A Course of
  Lectures}} (\bibinfo{publisher}{Academic Press}, \bibinfo{year}{2002}).

\bibitem[{\citenamefont{Bryngelson and Wolynes}(1987)}]{BW}
\bibinfo{author}{\bibfnamefont{J.~D.} \bibnamefont{Bryngelson}}
  \bibnamefont{and} \bibinfo{author}{\bibfnamefont{P.~G.}
  \bibnamefont{Wolynes}}, \bibinfo{journal}{Proc. Natl. Acad. Sci. U.S.A}
  \textbf{\bibinfo{volume}{84}}, \bibinfo{pages}{7524} (\bibinfo{year}{1987}).

\bibitem[{\citenamefont{Shakhnovich and Gutin}(1989)}]{SG_IndependentInt}
\bibinfo{author}{\bibfnamefont{E.~I.} \bibnamefont{Shakhnovich}}
  \bibnamefont{and} \bibinfo{author}{\bibfnamefont{A.~M.} \bibnamefont{Gutin}},
  \bibinfo{journal}{Biophys. Chem} \textbf{\bibinfo{volume}{34}},
  \bibinfo{pages}{187} (\bibinfo{year}{1989}).

\bibitem[{\citenamefont{Derrida}(1980)}]{REM_Derrida}
\bibinfo{author}{\bibfnamefont{B.}~\bibnamefont{Derrida}},
  \bibinfo{journal}{Phys. Rev. Lett} \textbf{\bibinfo{volume}{45}},
  \bibinfo{pages}{79} (\bibinfo{year}{1980}).

\bibitem[{\citenamefont{Sfatos and Shakhnovich}(1997)}]{Sfatos}
\bibinfo{author}{\bibfnamefont{C.~D.} \bibnamefont{Sfatos}} \bibnamefont{and}
  \bibinfo{author}{\bibfnamefont{E.~I.} \bibnamefont{Shakhnovich}},
  \bibinfo{journal}{Phys. Rep} \textbf{\bibinfo{volume}{288}},
  \bibinfo{pages}{77} (\bibinfo{year}{1997}).

\bibitem[{\citenamefont{Oliveberg and
  Shakhnovich}(2006)}]{Shakhnovich_review_2006}
\bibinfo{author}{\bibfnamefont{M.}~\bibnamefont{Oliveberg}} \bibnamefont{and}
  \bibinfo{author}{\bibfnamefont{E.~I.} \bibnamefont{Shakhnovich}},
  \bibinfo{journal}{Curr. Opin. Struct. Biol} \textbf{\bibinfo{volume}{16}},
  \bibinfo{pages}{68} (\bibinfo{year}{2006}).

\bibitem[{\citenamefont{Geissler et~al.}(2004)\citenamefont{Geissler,
  Shakhnovich, and Grosberg}}]{Solvation_random}
\bibinfo{author}{\bibfnamefont{P.~L.} \bibnamefont{Geissler}},
  \bibinfo{author}{\bibfnamefont{E.~I.} \bibnamefont{Shakhnovich}},
  \bibnamefont{and} \bibinfo{author}{\bibfnamefont{A.~Y.}
  \bibnamefont{Grosberg}}, \bibinfo{journal}{Phys. Rev. E}
  \textbf{\bibinfo{volume}{70}}, \bibinfo{pages}{021802}
  (\bibinfo{year}{2004}).

\bibitem[{\citenamefont{Khokhlov and Khalatur}(1999)}]{Khokhlov_PRL}
\bibinfo{author}{\bibfnamefont{A.~R.} \bibnamefont{Khokhlov}} \bibnamefont{and}
  \bibinfo{author}{\bibfnamefont{P.~G.} \bibnamefont{Khalatur}},
  \bibinfo{journal}{Phys. Rev. Lett} \textbf{\bibinfo{volume}{82}},
  \bibinfo{pages}{3456} (\bibinfo{year}{1999}).

\bibitem[{\citenamefont{Khokhlov and Khalatur}(2004)}]{Khokhlov_COSSMS}
\bibinfo{author}{\bibfnamefont{A.~R.} \bibnamefont{Khokhlov}} \bibnamefont{and}
  \bibinfo{author}{\bibfnamefont{P.~G.} \bibnamefont{Khalatur}},
  \bibinfo{journal}{Curr. Opin. Solid. State. Mater. Sci}
  \textbf{\bibinfo{volume}{8}}, \bibinfo{pages}{3} (\bibinfo{year}{2004}).

\bibitem[{\citenamefont{Pande et~al.}(1997)\citenamefont{Pande, Grosberg, and
  Tanaka}}]{BJ}
\bibinfo{author}{\bibfnamefont{V.~S.} \bibnamefont{Pande}},
  \bibinfo{author}{\bibfnamefont{A.~Y.} \bibnamefont{Grosberg}},
  \bibnamefont{and} \bibinfo{author}{\bibfnamefont{T.}~\bibnamefont{Tanaka}},
  \bibinfo{journal}{Biophys. J} \textbf{\bibinfo{volume}{73}},
  \bibinfo{pages}{3192} (\bibinfo{year}{1997}).

\bibitem[{\citenamefont{Li et~al.}(1996)\citenamefont{Li, Helling, Tang, and
  Wingreen}}]{Wingreen}
\bibinfo{author}{\bibfnamefont{H.}~\bibnamefont{Li}},
  \bibinfo{author}{\bibfnamefont{R.}~\bibnamefont{Helling}},
  \bibinfo{author}{\bibfnamefont{C.}~\bibnamefont{Tang}}, \bibnamefont{and}
  \bibinfo{author}{\bibfnamefont{N.}~\bibnamefont{Wingreen}},
  \bibinfo{journal}{Science} \textbf{\bibinfo{volume}{273}},
  \bibinfo{pages}{666} (\bibinfo{year}{1996}).

\bibitem[{\citenamefont{Reagan and Degrado}(1989)}]{Degrado}
\bibinfo{author}{\bibfnamefont{L.}~\bibnamefont{Reagan}} \bibnamefont{and}
  \bibinfo{author}{\bibfnamefont{W.~F.} \bibnamefont{Degrado}},
  \bibinfo{journal}{Science} \textbf{\bibinfo{volume}{241}},
  \bibinfo{pages}{976} (\bibinfo{year}{1989}).

\bibitem[{\citenamefont{Amatori et~al.}(2005)\citenamefont{Amatori, Tiana,
  F-Borg, Trovato, and Broglia}}]{Broglia1}
\bibinfo{author}{\bibfnamefont{A.}~\bibnamefont{Amatori}},
  \bibinfo{author}{\bibfnamefont{G.}~\bibnamefont{Tiana}},
  \bibinfo{author}{\bibfnamefont{J.}~\bibnamefont{F-Borg}},
  \bibinfo{author}{\bibfnamefont{A.}~\bibnamefont{Trovato}}, \bibnamefont{and}
  \bibinfo{author}{\bibfnamefont{R.~A.} \bibnamefont{Broglia}},
  \bibinfo{journal}{J. Chem. Phys} \textbf{\bibinfo{volume}{123}},
  \bibinfo{pages}{054904} (\bibinfo{year}{2005}).

\bibitem[{\citenamefont{Amatori et~al.}(2006)\citenamefont{Amatori, F-Borg,
  Tiana, and Broglia}}]{Broglia2}
\bibinfo{author}{\bibfnamefont{A.}~\bibnamefont{Amatori}},
  \bibinfo{author}{\bibfnamefont{J.}~\bibnamefont{F-Borg}},
  \bibinfo{author}{\bibfnamefont{G.}~\bibnamefont{Tiana}}, \bibnamefont{and}
  \bibinfo{author}{\bibfnamefont{R.~A.} \bibnamefont{Broglia}},
  \bibinfo{journal}{Phys. Rev. E} \textbf{\bibinfo{volume}{73}},
  \bibinfo{pages}{061905} (\bibinfo{year}{2006}).

\bibitem[{\citenamefont{Pande et~al.}(2000)\citenamefont{Pande, Grosberg, and
  Tanaka}}]{RMP}
\bibinfo{author}{\bibfnamefont{V.~S.} \bibnamefont{Pande}},
  \bibinfo{author}{\bibfnamefont{A.~Y.} \bibnamefont{Grosberg}},
  \bibnamefont{and} \bibinfo{author}{\bibfnamefont{T.}~\bibnamefont{Tanaka}},
  \bibinfo{journal}{Rev. Mod. Phys} \textbf{\bibinfo{volume}{72}},
  \bibinfo{pages}{259} (\bibinfo{year}{2000}).

\bibitem[{\citenamefont{Bouchaud and Mezard}(1997)}]{Bouchaud_Mezard}
\bibinfo{author}{\bibfnamefont{J.}~\bibnamefont{Bouchaud}} \bibnamefont{and}
  \bibinfo{author}{\bibfnamefont{M.}~\bibnamefont{Mezard}},
  \bibinfo{journal}{J. Phys. A: Math. Gen} \textbf{\bibinfo{volume}{30}},
  \bibinfo{pages}{7997} (\bibinfo{year}{1997}).

\bibitem[{\citenamefont{Dill and the others}(1995)\citenamefont{Dill
  et~al.}}]{Dill}
\bibinfo{author}{\bibfnamefont{K.~A.} \bibnamefont{Dill}} \bibnamefont{et~al.},
  \bibinfo{journal}{Protein Sci} \textbf{\bibinfo{volume}{4}},
  \bibinfo{pages}{561} (\bibinfo{year}{1995}).

\bibitem[{\citenamefont{Wong and Frishman}(2006)}]{Designability}
\bibinfo{author}{\bibfnamefont{P.}~\bibnamefont{Wong}} \bibnamefont{and}
  \bibinfo{author}{\bibfnamefont{D.}~\bibnamefont{Frishman}},
  \bibinfo{journal}{PLoS. Comput. Biol} \textbf{\bibinfo{volume}{2}},
  \bibinfo{pages}{392} (\bibinfo{year}{2006}).

\end{thebibliography}

\end{document}